\documentclass[aps,PRB,showpacs,groupedaddress,twocolumn]{revtex4}
\usepackage{graphicx}
\usepackage{amsmath}
\usepackage{subfigure}

\begin{document}

\title{Analytical model of the enhanced light transmission through
subwavelength metal slits: Green's function formalism versus
Rayleigh's expansion}

\author{S.\ V.\ Kukhlevsky$^a$, M.\ Mechler$^b$, O.\ Samek$^c$, K.\ Janssens$^d$}
\address{$^a$Department of Physics, Faculty of Natural Sciences,~University of
P\'ecs, Ifj\'us\'ag u.~6, P\'ecs 7624, Hungary\\
$^b$South-Trans-Danubian Cooperative Research Centre, University
of P\'ecs, Ifj\'us\'ag u.~6, P\'ecs 7624, Hungary\\
$^c$Institute of Spectrochemistry and Applied Spectroscopy,
Bunsen-Kirchhoff-Str.~11, D-44139 Dortmund, Germany\\
$^d$Department of Chemistry, University of Antwerp,
Universiteitsplein 1, B-2610 Antwerp, Belgium}

\begin{abstract}
We present an analytical model of the resonantly enhanced
transmission of light through a subwavelength nm-size slit in a
thick metal film. The simple formulae for the transmitted
electromagnetic fields and the transmission coefficient are
derived by using the narrow-slit approximation and the Green's
function formalism for the solution of Maxwell's equations. The
resonance wavelengths are in agreement with the semi-analytical
model [Y. Takakura, Phys. Rev. Lett. \textbf{86}, 5601 (2001)],
which solves the wave equations by using the Rayleigh field
expansion. Our formulae, however, show great resonant enhancement
of a transmitted wave, while the Rayleigh expansion model predicts
attenuation. The difference is attributed to the near-field
subwavelength diffraction, which is not considered by the
Rayleigh-like expansion models. 
\end{abstract}

\maketitle
\section{Introduction}
\label{sec:1} The near-field localization and resonantly enhanced
transmission of light by metal subwavelength nanostructures, such
as a single aperture, a grating of apertures and an aperture
surrounded by grooves, attract increasing interest of researchers.
The recent studies
\cite{Taka,Yang,Gar3,Thom,Brav,Lind,Suck,Xie,Kuk2,Liu} have
pointed out that the localization and enhanced (anomalous)
transmission of light by a grating of holes or slits can be better
understood by elucidating the optical properties of a single
subwavelength slit. Along this direction, it was already
demonstrated that a TM-polarized light wave can be localized in
the near-field subwavelength zone of a metal slit and
simultaneously enhanced by a factor of about
10-10$^3$~\cite{Yang,Gar3,Thom,Brav,Lind,Suck,Xie,Kuk2,Liu,Neer,Harr,Betz1}.
The general analysis and interpretation of these results, however,
are very complicated because the studies are based mainly on
purely numerical computer models. Recently, a simple
semi-analytical model of the light transmission by a subwavelength
metal slit was developed \cite{Taka}. The study clearly showed
that the transmission coefficient versus wavelength possesses a
Fabry-Perot-like resonant behavior. Unfortunately, the model
\cite{Taka} predicts transmission peaks with very low magnitude
(attenuation), while the experiments and computer models
~\cite{Yang,Gar3,Thom,Brav,Lind,Suck,Xie,Kuk2,Liu} demonstrate the
great resonant enhancement of a transmitted wave.

In this paper, we present an analytical model of the resonantly
enhanced transmission of light through a subwavelength nm-size
metal slit. The simple formulae for the transmitted
electromagnetic fields and the transmission coefficient are
derived by using the narrow-slit approximation and the Green's
function formalism for the solution of Maxwell's equations. The
article is organized as follows. The model and formulae are
described in section~\ref{sec:2}. In section~\ref{sec:3}, the
model is compared with the results of the semi-analytical model
\cite{Taka}. We show that the resonance wavelengths determined by
our formulae are in agreement with the model~\cite{Taka}, which
solves the wave equations by using the Rayleigh field expansion.
The formulae, however, indicate great resonant enhancement of a
transmitted wave, while the Rayleigh expansion model~\cite{Taka}
predicts attenuation. The summary and conclusion are given in
section~\ref{sec:4}.

\section{Analytical model based on Green's function formalism}
\label{sec:2}

\paragraph{Theoretical background}
Let us consider the scattering of TM-polarized light by a
subwavelength slit in a thick metallic film of perfect
conductivity. From the latter metal property it follows that
surface plasmons do not exist in the film. Such a metal is
described by the classic Drude model for which the plasmon
frequency tends towards infinity. We follow the Neerhoff and Mur
theoretical development based on the Green's function formalism for
the solution of Maxwell's equations \cite{Neer,Harr,Betz1}.

\begin{figure}[tb]
\begin{center}
\includegraphics[width=0.8\columnwidth]{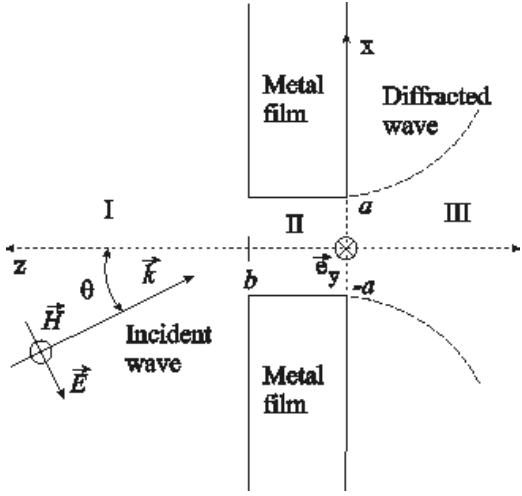}
\end{center}
\caption{Schematic diagram of light scattering by a subwavelength
slit (waveguide) in a thick metallic film.} \label{fig:fig1}
\end{figure}

The schematic diagram of the light scattering is shown in
Fig.~\ref{fig:fig1}.
The transmission of a plane wave through a subwavelength
($\lambda>2a$) slit of width $2a$ in a metal film of thickness $b$
is considered. In region I, the incident wave propagates in the
($x-z$) plane at an angle $\Theta$ with respect to the $z$ axis.
The magnetic field of the wave is assumed to be time harmonic with
the frequency $\omega=kc$ and both polarized and constant in the $y$
direction:
\begin{eqnarray}
{\vec{H}}(x,y,z,t)=U(x,z){\exp}(-i\omega{t}){\vec{e}}_y.
\end{eqnarray}
The electric field of the wave is found from the scalar field
$U(x,z)$ using Maxwell's equations. Thus, the electromagnetic
field is determined by a single scalar field $U(x,z)$. In the
regions I, II and III, the field is represented by $U_1(x,z)$,
$U_2(x,z)$ and $U_3(x,z)$, respectively. The field satisfies
the Helmholtz equation $(\nabla^2+k_j^2)U_j=0$, where $j=1,2,3$.
In region I, the field $U_1(x,z)$ is decomposed into
three components, $U_1(x,z)=U^i(x,z)+U^r(x,z)+U^d(x,z)$, each of
which satisfies the Helmholtz equation. The incident field
$U^i(x,z)=\exp[ik_1(x\sin\Theta-z\cos\Theta)$] is assumed to
be a plane wave of unit amplitude. $U^r(x,z)=U^i(x,2b-z)$ denotes
the field that would be reflected if there were no slit in the
film. $U^d$ describes the field diffracted by the slit into region
I. To find the field, the 2-dimensional Green's theorem is applied
with one function given by $U(x,z)$ and the other by a
conventional Green's function:
\begin{eqnarray}
(\nabla^2+k_j^2)G_j=-\delta(x-x',z-z'),
\end{eqnarray}
where $(x,z)$ refers to a field point of interest; $x'$ and $z'$
are integration variables, $j=1,2,3$. Since $U_j$ satisfies the
Helmholtz equation, Green's theorem reduces to
\begin{eqnarray}
U(x,z)=\int_{Boundary}(G\partial_n U-U\partial_n G)dS,
\end{eqnarray}
where $n$ is the normal vector at the boundary surface. The unknown fields $U^d(x,z)$, $U_3(x,z)$ and $U_2(x,z)$
respectively for the regions ($b<z<\infty$), ($-\infty<z<0$) and
($|x|<a$, $0<z<b$) are found using the reduced Green's theorem and
the standard boundary conditions for a perfectly conducting film
\begin{eqnarray}
U^d(x,z)&=&-\frac{\epsilon_1}{\epsilon_2}\int_{-a}^a G_1(x,z;x',b)DU_b(x')dx',\\
U_3(x,z)&=&\frac{\epsilon_3}{\epsilon_2}\int_{-a}^a G_3(x,z;x',0)DU_0(x')dx',\\
U_2(x,z)&=&-\int_{-a}^a [G_2(x,z;x',0)DU_0(x')\nonumber\\
&&-U_0(x')\partial_{z'}G_2(x,z;x',z')|_{z\rightarrow{0^+}}]dx'\nonumber\\
&&+\int_{-a}^{a}[G_2(x,z;x',b)DU_b(x')\nonumber\\
&&-U_b(x'){\partial}_{z'}G_2(x,z;x',z')|_{z\rightarrow{b^-}}]dx',
\end{eqnarray}
where the boundary fields are defined by
\begin{subequations}
\begin{eqnarray}
U_0(x)&\equiv& U_2(x,z)|_{z\rightarrow{0^+}},\label{eq:bcond1}\\
DU_0(x)&\equiv& {\partial}_zU_2(x,z)|_{z\rightarrow{0^+}},\label{eq:bcond2}\\
U_b(x)&\equiv& U_2(x,z)|_{z\rightarrow{b^-}},\label{eq:bcond3}\\
DU_b(x)&\equiv& {\partial}_zU_2(x,z)|_{z\rightarrow{b^-}}.\label{eq:bcond4}
\end{eqnarray}
\end{subequations}
In regions I and III the Green's functions are given by
\begin{eqnarray}
G_1(x,z;x',z')={\frac{i}{4}}[H_0^{(1)}(k_1R)+H_0^{(1)}(k_1R')],\\
G_3(x,z;x',z')={\frac{i}{4}}[H_0^{(1)}(k_3R)+H_0^{(1)}(k_3R'')],
\end{eqnarray}
with $R=[(x-x')^2+(z-z')^2]^{1/2}$,
$R'=[(x-x')^2+(z+z'-2b)^2]^{1/2}$, and
$R''=[(x-x')^2+(z+z')^2]^{1/2}$. Here, $H_0^{(1)}$ is the Hankel
function. Inside the waveguide (region II), the method of images
can be used; the Green's function is given by the waveguide
multimode ($m = {\infty}$) expansion
\begin{eqnarray}
G_2(x,z\lefteqn{;x',z')={\frac{i}{4a\gamma_0}}\exp(i\gamma_0|z-z'|)}\nonumber\\
&&+\frac{i}{2a}\sum_{m=1}^{\infty} \gamma_m^{-1}\exp(i{\gamma}{_m}|z-z'|)\nonumber\\
&&\times\cos\frac{m{\pi}(x'+a)}{2a}\cos\frac{m{\pi}(x+a)}{2a},
\end{eqnarray}
where $\gamma_m=[k_2^2-(m{\pi}/2a)^2]^{1/2}.$ The field can be
found once the four unknown functions in Eqs. \ref{eq:bcond1}-\ref{eq:bcond4} have been
determined. The functions are determined by the four integral
equations \cite{Betz1}:
\begin{eqnarray}
2U_b^i(x)-U_b(x)={\frac{\epsilon_1}{\epsilon_2}}\int_{-a}^{a}G_1(x,b;x',b)DU_b(x')dx',&&\label{eq:inteq1}\\
U_0(x)={\frac{\epsilon_3}{\epsilon_2}}\int_{-a}^{a}G_3(x,0;x',0)DU_0(x')dx',&&\label{eq:inteq2}
\end{eqnarray}
\begin{eqnarray}
{\frac{1}{2}}U_b(x)&=&-\int_{-a}^{a}[G_2(x,b;x',0)DU_0(x')\nonumber\\
&&-U_0(x'){\partial}_{z'}G_2(x,b;x',z')|_{z\rightarrow{0^+}}]dx'\nonumber\\
&&+\int_{-a}^{a}G_2(x,b;x',b)DU_b(x')dx',\label{eq:inteq3}\\
{\frac{1}{2}}U_0(x)&=&\int_{-a}^{a}[G_2(x,0;x',b)DU_b(x')\nonumber\\
&&-U_b(x'){\partial}_{z'}G_2(x,0;x',z')|_{z'\rightarrow{b^-}}]dx'\nonumber\\
&&-\int_{-a}^{a}G_2(x,0;x',0)DU_0(x')dx',\label{eq:inteq4}
\end{eqnarray}
where $|x|<a$, and
$U_b^i(x)=\exp[ik_1(x\sin{\Theta}-b\cos{\Theta})]$. A set of the
four coupled integral equations can be solved numerically by dividing
each integral in the expressions
(\ref{eq:inteq1}-\ref{eq:inteq4}) into $N$ subintervals of width $2a/N$ (for more details,
see \cite{Betz1}). The four boundary functions are found through
the following matrix equation:
\begin{small}
\begin{eqnarray}
\label{eq:matrix}
\begin{bmatrix}
\bf{D^{II}S^{III}-R^{II}}&\bf{S^{II}+\frac{1}{2}S^{I}}\\
\bf{S^{II}+\frac{1}{2}S^{III}}&\bf{D^{II}S^{I}-R^{II}}
\end{bmatrix}
\begin{bmatrix}
DU_0(x)\\
DU_b(x)
\end{bmatrix}
=
\begin{bmatrix}
U_b^i(x)\\
2\mathbf{D^{II}}U_b^i(x)
\end{bmatrix},\quad
\end{eqnarray}
\end{small}
where
\begin{subequations}
\begin{eqnarray}
R_{kj}^{II}&=&\frac{i}{2N\gamma_0}\exp(i\gamma_0b)\nonumber\\
&&+\frac{i}{N}\sum_{m=1}^{\infty}\frac{1}{\gamma_m}T_m^{j,k}\exp(i\gamma_mb),\\
D_{kj}^{II}&=&\frac{1}{2N}\exp(i\gamma_0b)\nonumber\\
&&+\frac{1}{N}\sum_{m=1}^{\infty}T_m^{j,k}\exp(i\gamma_mb),\\
S_{kj}^{II}&=&\frac{i}{2N\gamma_0}+\frac{i}{N}\sum_{m=1}^{\infty}\frac{1}{\gamma_m}T_m^{j,k},\\
T_m^{j,k}&=&\frac{2N}{m\pi}\cos\left[\frac{m\pi(j-1/2)}{N}\right]\nonumber\\
&&\times\cos\left[\frac{m\pi(k-1/2)}{N}\right]\sin\left[\frac{m\pi}{2N}\right].
\end{eqnarray}
\end{subequations}

\paragraph{The analytical model and formulae}
Let us now present an approximate analytical solution of the
scattering problem. For a subwavelength slit, the narrow-slit
condition $2a \ll \lambda$ is a very good approximation. In such a
case, it can be easily demonstrated that an accurate solution
exists for the above matrix equation in the case of $N = 1$, which
is valid at $|z| > 2a/N$ (see, ref.~\cite{Betz1}). We have derived
the simple formulae for the transmitted electromagnetic field and
the transmission coefficient using the narrow-slit approximation.
After simple calculation using the above mentioned conditions,
the analytical formulae for the magnetic ${\vec{H}}(x,z)$ and
electric ${\vec{E}}(x,z)$ fields in region III can be presented as
\begin{align}
&\vec{H}(x,z)=i{a}H_0^{(1)}(k\sqrt{x^2+z^2})(D{{U}}_0)_{1}{\vec{e}}_y,\label{eq:hvec-veg}\\
&E_{x}(x,z)=-a\frac{z}{\sqrt{x^2+z^2}}H_1^{(1)}(k\sqrt{x^2+z^2})(D{U}_0)_1,\label{eq:ex-veg}\\
&E_{y}(x,z)=0\label{eq:ey-veg}\\
\intertext{and}
&E_{z}(x,z)=a{\frac{x}{\sqrt{x^2+z^2}}}H_1^{(1)}(k\sqrt{x^2+z^2})(D{{U}}_0)_{1},\label{eq:ez-veg}
\end{align}
where
\begin{eqnarray}
\label{sz:D:def}
(D{{U}}_0)_{1}=\frac{4}{ik}\frac{\exp(ikb-ikb\cos\theta)}
{[\exp(ikb)\cdot(A-{i}k)]^2-(A+{i}k)^2}
\end{eqnarray}
and
\begin{eqnarray}
\label{sz:A:def}
A=ia\left[H_0^{(1)}\left(ka\right)+\left(\frac{\pi}{2}\right)[\vec{F}_0\left(ka\right)H_1^{(1)}\left(ka\right)\right.\nonumber\\
\left.-\vec{F}_1\left(ka\right)H_0^{(1)}\left(ka\right)]\right].
\end{eqnarray}
Here, $H_0^{(1)}(ka)$ and $H_1^{(1)}(ka)$ are the Hankel
functions, and $\vec{F}_0(ka)$ and $\vec{F}_1(ka)$ are the Struve
functions; for the sake of simplicity, we presented the formulae
for the slit placed in vacuum. The transmission coefficient is
determined by the time averaged Poynting vector (energy flux)
$\vec S$ of the electromagnetic field. The vector is calculated as
$\vec{S}=(\vec{E}\times\vec{H}^*+\vec{E}^*\times\vec{H})$. The
transmission coefficient $T$ is defined as the integrated
transmitted flux $S^3(z)$ divided by the integrated incident flux
$S^i$. The flux $S^3(z)$ is integrated at the distance $z$
($z>{2a}$) from $x_{min}$ = -$\infty$ to $x_{max}$ = $\infty$, and
the flux $S^i$ is integrated over the slit width $2a$ at the slit
entrance. For the transmission coefficient we have obtained the
following simple analytical formula:
\begin{eqnarray}
T(\lambda,a,b)=\frac{a}{k\cos\theta}\cdot\left[(\mathrm{Re}(D{{U}}_0)_{1})^2
+(\mathrm{Im}(D{{U}}_0)_{1})^2\right].\quad\label{eq:trans-simple}
\end{eqnarray}
Notice, that in agreement with the energy conservation condition,
the transmission coefficient does not depend on the distance $z$.
This means that the near-field and far-field transmission coefficients are
of equal value, $T(2a<{z}<{\lambda})=T(z>\lambda)$.

\section{Comparison with model based on Rayleigh's field expansion}
\label{sec:3}

The analytical formulae (\ref{eq:hvec-veg}-\ref{eq:trans-simple}) for the transmitted
electromagnetic field and the transmission coefficient have been
derived by using the narrow-slit ($2a\ll \lambda$) approximation
and the Green's function formalism for the solution of Maxwell's
equations. The main objective of our analysis is to test the
accuracy and range of validity of the formulae. We have calculated
the magnetic ${\vec{H}}(x,z)$ and electric ${\vec{E}}(x,z)$ fields
of the transmitted wave and the transmission coefficient $T$ for
different parameters of the slit. As an example, Figs.~2-5 show
the numerical results for the incident angle $\Theta$ equal to
zero. Figure~2(a) shows the field distributions $|H_y(x,z)|$, $|E_x(x,z)|$ and $|E_z(x,z)|$
calculated by the analytical formulae (\ref{eq:hvec-veg}-\ref{eq:ez-veg}) for the slit
width $2a=150$~nm and the screen thickness $b=150$~nm. The field distributions
calculated at the distance $|z|=2a$ by the numerical evaluation~\cite{Betz1} of the integral equations
(\ref{eq:inteq1}-\ref{eq:inteq4}) are shown in Fig. 2(b) for the comparison.
\begin{figure}[tb]
\begin{center}
\subfigure[]{\includegraphics[keepaspectratio,width=0.8\columnwidth]{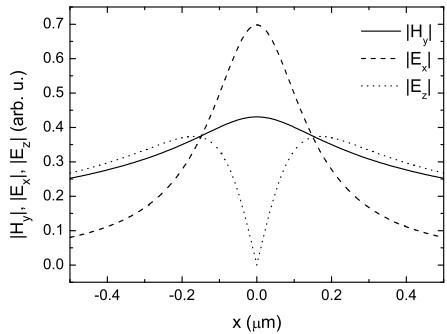}}
\subfigure[]{\includegraphics[keepaspectratio,width=0.8\columnwidth]{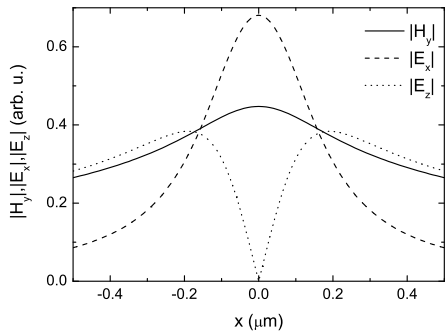}}
\end{center}
\caption{The field distributions $|H_y|$, $|E_x|$ and $|E_z|$ of the transmitted wave with $\lambda=2500$~nm
calculated at $|z|=2a$ for the slit width $2a=150$~nm and the screen thickness $b=150$~nm:
(a) analytical model and (b) rigorous numerical model.} \label{fig:fig2-1}
\end{figure}
\begin{figure}[tb]
\begin{center}
\subfigure[]{\includegraphics[keepaspectratio,width=0.8\columnwidth]{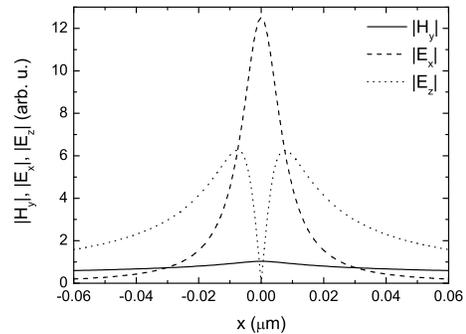}}
\subfigure[]{\includegraphics[keepaspectratio,width=0.8\columnwidth]{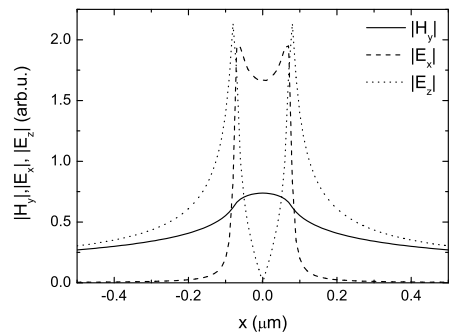}}
\end{center}
\caption{The field distributions $|H_y|$, $|E_x|$ and $|E_z|$ of the transmitted wave with $\lambda=2500$~nm
calculated at $|z|=0.1a$ for the slit width $2a=150$~nm and the screen thickness $b=150$~nm:
(a) analytical model and (b) rigorous numerical model.} \label{fig:fig2-2}
\end{figure}
We notice that the distributions are practically undistinguishable at $|z|=2a$. The calculations show that
the accuracy of the analytical formulaes increases with further increasing the distance $|z|$.
The field components $H_y$ and $E_x$ are collimated to approximately the aperture width $2a$ (see, Fig.~2).
Thus, the energy flux $S_z$ of a wave passing through a subwavelength slit can be used to provide a
subwavelength image with the optical resolution of about $2a$. We calculated the magnetic ${\vec{H}}(x,z)$
and electric ${\vec{E}}(x,z)$ fields of the transmitted wave also in the region ($|z|<2a$) by using the
analytical formula (Fig. 3(a)) and by the numerical evaluation~\cite{Betz1} of the integral equations
(\ref{eq:inteq1}-\ref{eq:inteq4}) (Fig. 3(b)). The results show that the accuracy of the formulae
(\ref{eq:hvec-veg}-\ref{eq:ez-veg}) decreases with decreasing the value $|z|$ in the region ($|z|<2a$).

The transmission of light by the slit is a process that depends on
the wavelength. Due to the dispersion, the amplitude of a wave
does change under propagation through the slit. This leads to
attenuation or amplification of the wave intensity. In our
analytical model, the transmission coefficient $T(\lambda,a,b)$ is
described by the analytical formula (\ref{eq:trans-simple}). We now check the
consistency of the results by comparing the transmission
coefficient $T(\lambda)$ calculated by using the semi-analytical
model~\cite{Taka} with those obtained by the formula (\ref{eq:trans-simple}).
Analysis of Eq.~\ref{eq:trans-simple} indicates the slit-wave interaction behavior that
is similar to those of a Fabry-Perot resonator. The minimum
thickness of the screen is required to get the waveguide resonance
inside the slit at a given wavelength. Figure~4 shows the
transmission coefficient (curve~A)
\begin{figure}[tb]
\begin{center}
\includegraphics[keepaspectratio,width=\columnwidth]{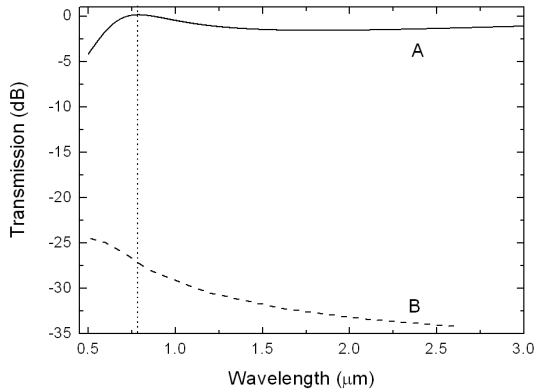}
\end{center}
\caption{The transmission (curve A) as a function of the
wavelength calculated by using the analytical formula (\ref{eq:trans-simple})
for the slit width $2a=150$~nm and the screen thickness $b=150$~nm.
The curve B shows the transmission calculated in the study~\cite{Taka}.
The dotted vertical line corresponds the resonant wavelength given by the analytical formula.}
\label{fig:fig3}
\end{figure}
as a function of the wavelength $\lambda$ calculated by using the
formula (\ref{eq:trans-simple}) for the slit width $2a=150$~nm and the screen thickness ($b=150$~nm) smaller than the wavelength $\lambda$. The transmission coefficient (curve~B) calculated in the study~\cite{Taka} is presented in the figure for comparison. In contrast to the study~\cite{Taka}, the transmission (curve A) calculated by our formula exhibits resonance peak, while curve B demonstrates no resonances. In Fig.~4, the screen thickness has been chosen in such a way as to include one Fabry-Perot resonant wavelength $\lambda_{FP}=2b$. Notice that the resonance peak is red shifted in comparison with the respective Fabry-Perot wavelength $\lambda_{FP}=300$~nm. In Fig.~5, the slit includes several Fabry-Perot resonances at the wavelengths $\lambda_{FP}=2b/n$. Figure~5 shows the transmission coefficient as a
\begin{figure}[tb]
\begin{center}
\includegraphics[keepaspectratio,width=\columnwidth]{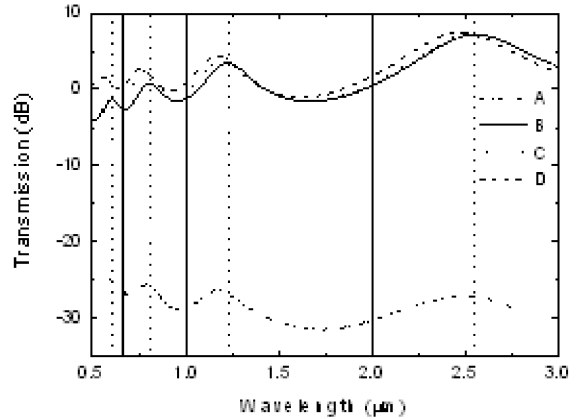}
\end{center}
\caption{The transmission as a function of the wavelength calculated for the slit width $2a=150$~nm and the screen thickness $b=1000$~nm. The curve A shows the transmission calculated in the study~\cite{Taka}. The curve B shows the transmission calculated by using the analytical formula (\ref{eq:trans-simple}). The transmission coefficients calculated by the numerical evaluation~\cite{Betz1} of the integral equations (\ref{eq:inteq1}-\ref{eq:inteq4}) for the far-field ($z=1$~mm) and near-field ($z=150$~nm) zones are shown by the curves C and D, respectively. The dotted vertical lines correspond the resonant wavelengths given by the analytical formula. The respective Fabry-Perot resonances are indicated by vertical solid lines.} \label{fig:fig4}
\end{figure}
function of the wavelength $\lambda$ for the slit width $2a=150$~nm and the screen
thickness ($b=1000$~nm) greater than the wavelength. The figure shows the transmission spectra calculated in the study~\cite{Taka} and the transmission coefficients calculated by our analytical formula and by the numerical evaluation~\cite{Betz1} of the integral equations (\ref{eq:inteq1}-\ref{eq:inteq4}) for the near-field ($z=150$~nm) and far-field ($z=1$~mm) zones. The transmission spectra indicates the slit-wave interaction behavior, which is similar to those of a Fabry-Perot resonator. The transmission resonance peaks, however, have a systematic shift from the Fabry-Perot wavelength $\lambda_{FP}=2b/n$ ($n=1,2,3 \dots$) towards longer wavelengths. We notice that the resonance wavelengths determined by our formulae are in agreement with the model~\cite{Taka}, which solves the wave equations by using the Rayleigh field expansion. The peak heights (curves B, C and D), however, are different from the results~\cite{Taka}. Our analytical formula predicts the great resonant enhancement of a transmitted wave (curve B), while the Rayleigh expansion model~\cite{Taka} indicates attenuation (curve A). Notice that the curves B, C and D are practically undistinguishable in the case of the narrow slit ($2a\ll\lambda$). The transmission coefficient $T$ calculated by the analutical formula (curve B) and the transmission coefficients calculated by the numerical evaluation~\cite{Betz1} of the integral equations (curves C and D) go above unity (i.~e.,~0 dB) at some resonant wavelengths (see, Fig. 5). Such behaviour of the transmission is strange from a point of view of the classical optics. Indeed, according to the classical geometrical ($\lambda{\ll}2a$) or Fresnel-Kirchhoff diffraction ($\lambda<2a$) optics, the light energy impinging on the slit opening only contributes to the transmitted energy. The screen completely absorbs the light outside the slit aperture. Consequently, the value $T$ cannot exceed unity. In the case of the subwavelength ($\lambda>2a$) slit in a thick ($b>\lambda/2$) metal screen, the screen does not absorb totally the light energy outside the slit aperture. At the appropriate resonant conditions, the system redistributes the electromagnetic energy in the intra-slit region and around the screen, such that more energy ($T>1$) is effectively transmitted compared to the energy impinging on the slit opening. The total energy of the system is conserved under the energy redistribution. In our plasmon-less model, the redistribution of light energy is caused by the boundary conditions imposed on the electromagnetic wave at the metal surface. In the plasmon-based models the light energy is redistributed due to the boundary conditions for both the electromagnetic and electron waves. It can be noted that the transmission coefficient ($T_{max}\sim \lambda/2{\pi}a$ at the resonance) increases with increasing the wavelength. For instance, in the experiment \cite{Yang} the coefficient $T\sim400$ was achieved in the microwave spectral region.

Analysis of the denominator of the formulae (\ref{sz:D:def}-\ref{eq:trans-simple}) shows that the transmission will exhibit the maximums around the Fabry-Perot wavelengths $\lambda_{FP}=2b/n$ ($n=1,2,3 \dots$). The shifts of the resonance wavelengths from the values $\lambda_{FP}=2b/n$
are caused by the wavelength dependent term in the denominator of
Eq.~\ref{eq:trans-simple} (see, Eqs.~\ref{sz:D:def}, \ref{sz:A:def}). The values of the resonant wavelengths and the shifts calculated using the analytical formula (\ref{eq:trans-simple}) are in good agreement with the values calculated using the analytical formula of the study ~\cite{Taka}. The formula (\ref{eq:trans-simple}), however, shows the great resonant enhancement of a transmitted wave, while the Rayleigh expansion model~\cite{Taka} predicts attenuation. The difference between the Rayleigh and Green-function models is attributed to the influence of the near-field subwavelength diffraction, which is not considered by the models based on the
Rayleigh field expansion. Indeed, in the Rayleigh-like expansion models the fields are found by matching the intra-slit field $U_2(x,z)$ with the external fields $U_1(x,z)$ and $U_3(x,z)$ at the slit entrance ($z=0$) and
exit ($z=b$). The field $U_2(x,z)$ is presented as a sum of the infinite number of waveguide modes, the cavity mode expansion. In the most cases, the
field $U_2(x,z)$ can be well approximated by a lowest order waveguide mode. The field $U_1(x,z)$ is given by a superposition of the infinite number of plane waves, the plane wave expansion. The similar plane wave expansion is used for the field $U_3(x,z)$. In the case of $\lambda<2a$, the near-field ($z\rightarrow{0^-}$ or $z\rightarrow{b^+}$) diffraction is weak. Consequently, the fields $U_1(x,z)$ and $U_3(x,z)$ can be considered as nearly plane waves. Such waves are well described by a few terms ($m=0,1,\dots ,M$) in the plane wave expansion. For instance, the classic Airy formula for the transmission coefficient is obtained by using the zero-order ($M=0$) approximation of the fields, $U_1(x,z)\sim U_1(z)$ and $U_3(x,z)\sim U_3(z)$. In the case of subwavelength ($\lambda>2a$) slits, the near-field diffraction is strong. The spatial distribution and energy of the fields $U_1(x,z)$ and $U_3(x,z)$ are very different from the plane waves. To obtain an accurate solution, one should use a great number of terms in the plane wave
expansion of the fields $U_1(x,z)$ and $U_3(x,z)$. Unfortunately, the Rayleigh-like field expansion models use only a few terms. Thus, the fields and transmission coefficient calculated by using such models can be considered as a very rough approximation. In the case of the Green's function model, the rigorous solutions for the fields $U_1(x,z)$, $U_2(x,z)$ and $U_3(x,z)$ are found by numerical evaluation of the integral equations (\ref{eq:inteq1}-\ref{eq:inteq4}). In the case of the narrow slits, we have found an accurate analytical solution of the integral equations, which yields the formulae (\ref{eq:hvec-veg}-\ref{eq:trans-simple}) for the electromagnetic fields and transmission coefficient.

\section {Summary and conclusion}
\label{sec:4}

We have presented an analytical model of the resonantly enhanced transmission of light through a subwavelength nm-size slit in a thick metal film. The simple formulae for the transmitted electromagnetic fields and the transmission coefficient were derived by using the narrow-slit approximation and the Green's function formalism for the solution of Maxwell's equations. The resonance wavelengths are in agreement with the semi-analytical model~\cite{Taka}, which solves the wave equations by using the Rayleigh field expansion. Our formulae, however, show great resonant enhancement of a transmitted wave, while the Rayleigh expansion model predicts attenuation. The transparent explanation of the difference between the results of the two models was presented. The difference is attributed to the near-field subwavelength diffraction, which is not considered by the Rayleigh-like expansion models. We believe that the presented analytical model gains insight into the physics of resonant transmission and localization of light by subwavelength nanoapertures in metal films.

\begin{acknowledgements}
This study was supported by the Fifth Framework of the European
Commission (Financial support from the EC for shared-cost RTD
actions: research and technological development projects,
demonstration projects and combined projects. Contract No
NG6RD-CT-2001-00602) and in part by the Hungarian Scientific
Research Foundation (OTKA, Contract Nos T046811 and M045644) and
the Hungarian R\& D Office (KPI, Contract No
GVOP-3.2.1.-2004-04-0166/3.0).
\end{acknowledgements}

\end{document}